\documentclass{Interspeech}

\interspeechcameraready

\title{Prosodically Enhanced Foreign Accent Simulation\\
by Discrete Token-based Resynthesis
Only with Native Speech Corpora}

\author[affiliation={1,2}]{Kentaro}{Onda}
\author[affiliation={3}]{Keisuke}{Imoto}
\author[affiliation={2}]{Satoru}{Fukayama}
\author[affiliation={1}]{Daisuke}{Saito}
\author[affiliation={1}]{Nobuaki}{Minematsu}

\affiliation{}{The University of Tokyo}{Japan}
\affiliation{}{National Institute of Advanced Industrial Science and Technology (AIST)}{Japan}
\affiliation{}{Kyoto University}{Japan}
\email{\{ondakentaro, mine\}@gavo.t.u-tokyo.ac.jp}

\usepackage{comment}
\usepackage[dvipdfmx]{graphicx}
\usepackage{url}
\usepackage{booktabs}
\usepackage{multirow}
\usepackage{xcolor}

\begin{document}

\maketitle

\begin{abstract}
  Recently, a method for synthesizing foreign-accented speech only with native speech data using discrete tokens obtained from self-supervised learning (SSL) models was proposed. Considering limited availability of accented speech data, this method is expected to make it much easier to simulate foreign accents. By using the synthesized accented speech as listening materials for humans or training data for automatic speech recognition (ASR), both of them will acquire higher robustness against foreign accents. However, the previous method has a fatal flaw that it cannot reproduce duration-related accents. Durational accents are commonly seen when L2 speakers, whose native language has syllable-timed or mora-timed rhythm, speak stress-timed languages, such as English. In this paper, we integrate duration modification to the previous method to simulate foreign accents more accurately. Experiments show that the proposed method successfully replicates durational accents seen in real L2 speech.
\end{abstract}

\section{Introduction}
\label{sec:intro}

In today's globalized world, it is common for people to speak languages other than their native ones. 
Furthermore, recent trends in foreign language education have emphasized intelligible pronunciation over native-like pronunciation, 
adopting the ``intelligibility principle" \cite{munro1995foreign, murphy2014intelligible, levis2020revisiting}. 
Particularly in the case of English as a lingua franca, 75\% of the 1.5 billion
speakers are non-native \cite{ethnologue}, and now
it is more common to encounter accented pronunciations often referred to as World Englishes \cite{kachru1992world}. 
These accents can be challenging for both human listeners \cite{McLaughlin2018} and machine listeners \cite{racial} to understand,
primarily due to the lack of exposure to accented speech.
Accented speech is rarely used as listening materials in education \cite{nakanishi}, 
and non-native speech corpora are significantly fewer compared to native speech ones \cite{mitigating, Klumpp2023SyntheticCD}. 
For ASR, solutions such as accent reduction as preprocessing \cite{modification} and data augmentation through accent conversion \cite{Klumpp2023SyntheticCD} have been proposed, 
but these methods generally rely on training with accented speech. 
This limits their applicability only to a few accents available in existing corpora, 
which are mostly ``X-accented English" leaving out accents in languages other than English.
Considering that there are lots of people who have to speak languages other than English as a second language,
this limitation is a significant issue for truly global communications.

In \cite{onda24_interspeech}, a method was proposed to synthesize foreign-accented speech using only native speech corpora, 
without using any accented speech data for training. 
In the context of cross-lingual voice conversion, in which only native speech data is also used for training,
some studies have reported that unwanted accentuation occurs in the converted speech as a by-effect \cite{zhou2023,zhang2024refxvccrosslingualvoiceconversion}.
To the best of our knowledge, however, \cite{onda24_interspeech} is the first study to intentionally simulate the conginitive mechanism of foreign accentuation and investigate the similarity between the synthesized and real accented speech.
The method utilizes discrete token-based speech resynthesis to simulate the actual cognitive process of foreign accentuation and
convert native speech of language A into 
its varient with the accent of language B. 
Since the method only uses native speech data of language B for training and does not depend on the language of input speech,
it can generate various combinations of ``B-accented A," including accents for which no actual speech data is available.
However, \cite{onda24_interspeech} was still a pilot study and had several limitations. 
One of the key issues mentioned in the paper was its inability to reproduce accents related to phoneme durations.
Duration control is a crucial factor in foreign accents and significantly affects the intelligibility of L2 speech \cite{duration, WINTERS2013486, TAJIMA19971}. 
For example, 
when native speakers of syllable-timed or mora-timed languages, such as Japanese, 
speak English, a stress-timed language,
vowel lengths tend to be equalized regardless of stress \cite{riney1993japanese, stress}.
Therefore, this paper aims to improve the conventional method 
by incorporating the replication of durational accents
using only native speech corpora again. 
Additionally, \cite{onda24_interspeech} focused solely on phonetic accents when evaluating the method by measuring the similarity 
between synthetic and real accented speech. Since suprasegmental features are also important factors of foreign accents \cite{kang_supra},
this study expands the evaluation to include prosodic features such as intonation, intensity, and duration.
A subjective evaluation by experts was also conducted to assess 
the overall naturalness of the synthesized accents.

\section{Simulation of foreign accentuation by discrete token-based resynthesis}
\label{sec:related}
In \cite{onda24_interspeech}, a method was proposed to synthesize foreign-accented speech using only native speech corpora by simulating
the mechanism of foreign accentuation influenced by the speaker's native language.
In this method, discrete tokens used in Generative Spoken Language Model (GSLM) \cite{lakhotia-etal-2021-generative} 
are interpreted as phonological representations \cite{Baddeley2007}, 
which are the results of the listeners' cognitive process of speech perception.
Accordingly, the discrete token-based speech resynthesis is regarded as 
a human-like repetition process involving perception and reconstruction of the input speech.
Foreign accents emerge when a speaker perceives foreign speech through the framework of their native language and reconstructs it using the articulatory patterns of their native language \cite{flege1995second}.
Therefore, by resynthesizing input speech from language A using an encoder and decoder trained on native speech of language B,
the output speech is expected to be ``B-accented A" speech.


\subsection{Generative Spoken Language Model}
\label{ssec:gslm}
GSLM \cite{lakhotia-etal-2021-generative} is a language model built solely on speech data, without any text.
By discretizing SSL representations, such as HuBERT \cite{hubert} and wav2vec 2.0 \cite{wav2vec}, 
using k-means clustering,
the resulting discrete tokens are treated as pseudo-text.
GSLM consists of three components: Speech-to-unit (S2u), unit-Language-Model (uLM), and unit-to-Speech (u2S).
S2u converts speech to discrete tokens (units), uLM predicts the next unit given the previous units,
and u2S converts the units back to speech by training an existing TTS model 
treating the units as input text. 
In \cite{lakhotia-etal-2021-generative}, Tacotron2 \cite{taco} was employed.
By cascading these three models, the system can predict and synthesize a continuation of the input speech.
If uLM is omitted, the system functions as a simple speech resynthesis model. 
In this study, only S2u and u2S are utilized as the encoder and decoder, respectively.
\subsection{Simulation of foreign accentuation}
\label{ssec:simulation}
\cite{onda24_interspeech} is a pilot study of foreign accent simulation 
through discrete token-based resynthesis using S2u and u2S of GSLM.
Segmental analysis
revealed that the synthesized accented speech
successfully reproduces the accentuation tendencies observed in real L2 speakers. 
However, the study explicitly stated that the method could not reproduce accents on duration, 
and the evaluation of the synthesized accent was limited to its segmental aspect.
Therefore, this study aims to enhance the method by replicating accents also on duration 
and evaluate the resultant accent from a prosodic perspective.

\section{Duration Modeling}
\label{sec:duration}
\begin{figure}[t]
  \centering
  \hspace*{-16mm}
  \includegraphics[width=1.3\linewidth]{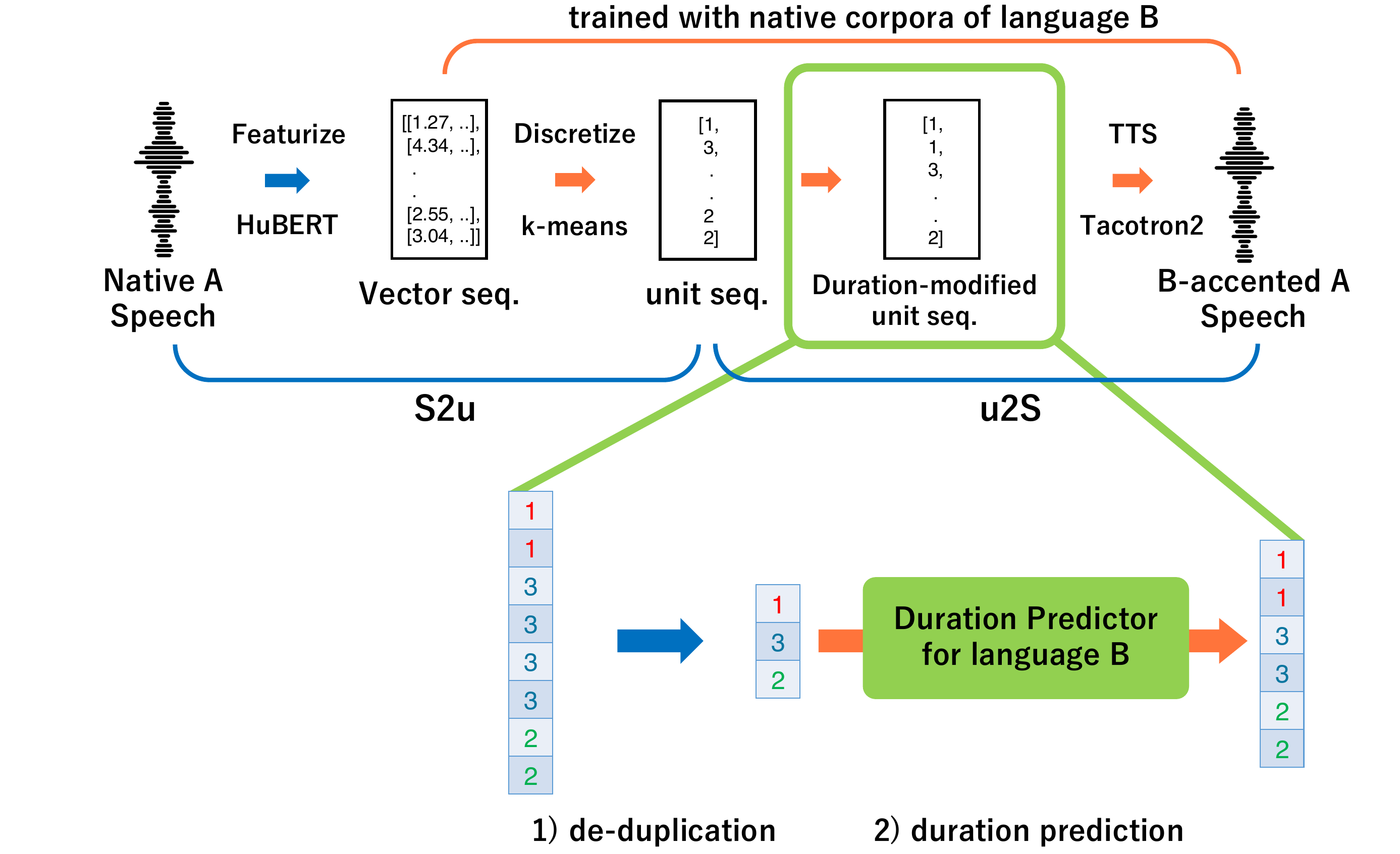}
  \caption{Foreign accent simulation with duration modeling}
  \label{fig:resynth}
  \vspace*{-4mm}
\end{figure}
As shown in Fig. \ref{fig:resynth}, we add a unit-based duration modification module to 
the original u2S framework of \cite{onda24_interspeech}.
Since the original units are the results of frame-by-frame discretization of input native speech,  
the temporal structure of the input is preserved.
As a result, the duration of the individual phonemes in output speech
remains nearly identical to that of the input speech, failing to reproduce accents on duration.
To address this issue, we introduce a module to appropriately modify the duration of each unit.
This module can also be trained with only native speech corpora, as the accentuation 
observed in real L2 speakers is caused by their L1 influence
rather than by imitation of other L2 speakers.
By modifying the unit sequence using this module before passing it to the TTS module, 
the output speech is expected to exhibit accents on duration as well.

The modification of the unit sequence consists of two steps: 1) de-duplication and 2) duration prediction.
If the duration predictor is trained on language B, 
it is expected to modify the duration of each unit to resemble that of language B, even when the input is in language A.
Here, we employed the same duration predictor architecture commonly used for phoneme length modeling 
in non-autoregressive TTS such as FastSpeech2 \cite{ren2021fastspeech} and VITS \cite{kim2021conditional}.
In TTS, phoneme sequences and the forced alignment results are used to train the duration predictor.
However, for unit-based modeling, the original frame-wise S2u output can be regarded as the ground truth.
Therefore, no alignment procedure is required, and the duration predictor can be trained 
by treating units as phoneme-like representations.
In this study, we employed the two-layer convolutional network from FastSpeech2 for duration prediction.
This method explicitly modifies the duration of each unit; 
however, an alternative approach is to perform only de-duplication and allow the decoder to 
implicitly predict the duration of each unit while converting the de-duplicated unit sequence into 
acoustic features.
We will evaluate both approaches in the following experiment.

\section{Experiment}
\label{sec:experiment}
\subsection{Experimental setup}
Using the same native speech corpora as \cite{onda24_interspeech}, we trained three modules: 
S2u, u2S, and a duration predictor for Japanese units.
Then, by inputting native English speech,
we synthesized Japanese-accented English, 
where durational accents are expected to appear \cite{riney1993japanese}.
The combination of JSUT \cite{sonobe2017jsut}, JVS \cite{takamichi2019jvs}, JKAC \cite{jkac}, and JMAC \cite{takamichi2022jmac} was used to train S2u, 
and only JSUT, a single-speaker corpus,
was used to train u2S and the duration predictor.
All the corpora used here consists entirely of Japanese speech read by native Japanese speakers, 
without any native English or Japanese-accented English speech.
Therefore, the same model can be used to synthesize any ``Japanese-accented A" speech, 
where A represents any language. However, for evaluation, we used English as A.
As an SSL model, we used the 6th layer of HuBERT-base \cite{hubert} following previous studies \cite{onda24_interspeech,lakhotia-etal-2021-generative}.

The training procedure is as follows:
1) Train S2u, k-means clustering of the SSL features, using the above corpora, 
2) Convert all speech samples in JSUT to units using the trained S2u,
3) Train the duration predictor and u2S using the converted units and original speech samples in JSUT.
As a reference, we also trained a model only with de-duplication.
In this case, the TTS module implicitly predicts the duration of each unit.
The original model from \cite{onda24_interspeech} was used as the baseline without any modification.
To investigate the effect of the number of unique units (the `k' in k-means), 
we trained the models with three different cluster sizes: 50, 200, 1000.


\subsection{English Read by Japanese corpus}
ERJ \cite{Minematsu2004DevelopmentOE} is a corpus of English read by Japanese speakers, which is used as the target accent in this study.
This corpus contains the same sets of sentences and words read aloud by both Japanese learners of English and native speakers of American English.
We used native English speech as input and compared the output with real Japanese English.
In this study, we use the phonetic sentence section of the ERJ corpus. 
This section consists of 460 phoneme-balanced sentences. 
Each sentence is read aloud by about 24 Japanese learners of English and 11 native English speakers. 
All 5046 utterances by native English speakers are used as input speech in the following experiment.

\subsection{Unit-based analysis}
\begin{table}[tb]
  \centering
  \caption{Mean and standard deviations of unit-based duration}
  \label{tab:unit}
  \vspace*{-1mm}
  \hspace*{-6mm}
  \resizebox{1.1\columnwidth}{!}{
    \begin{tabular}{lccccc}\toprule
      & \multicolumn{2}{c}{Native Japanese (rNJ)} & \multicolumn{2}{c}{American English (rAE)} & \multicolumn{1}{c}{Japanese English (rJE)} \\
      & baseline & dur-mod& \textbf{baseline}& \textbf{dur-mod}& baseline \\ \midrule
     50 & 1.79 / 1.14 & 1.78 / 1.13  & 2.10 / 1.65 & 1.89 / 1.37 & 2.17 / 1.62 \\
     200 & 1.49 / 0.82 & 1.46 / 0.80  & 1.70 / 1.23 & 1.44 / 0.82 &1.73 / 1.15  \\
     1000 & 1.26 / 0.58 & 1.16 / 0.50  & 1.45 / 0.97 & 1.18 / 0.53 &1.46 / 0.87   \\\bottomrule
    \end{tabular}
  }
  \vspace*{-3mm}
\end{table}
Before looking at the output speech, we analyzed the effectiveness of the duration modification module on the unit sequences.
Table \ref{tab:unit} shows the mean and standard deviation (SD) of the unit duration of the input (baseline) 
and output (dur-mod) of the duration modification module, expressed as mean/SD.
Unit duration is defined as the number of consecutive repetitions of the same unit.
For example, given the sequence (\textless2\textgreater, \textless2\textgreater, \textless1\textgreater, \textless2\textgreater, \textless3\textgreater, \textless3\textgreater),
the corresponding unit durations would be (2, 1, 1, 2).
We compared the cases where the input was real Native Japanese (rNJ), and real American English (rAE).
The baseline and dur-mod of the rAE (shown in bold) are the token sequences fed 
into the TTS module in the subsequent experiment 
to synthesize Japanese-accented English.
The baseline output of real Japanese English (rJE) is also shown as a reference, 
as it is expected to align with the temporal structure of accented speech produced by real Japanese speakers.
JSUT was used for rNJ, while ERJ was used for rAE and rJE.


For rNJ, the mean and SD before and after the duration modification (1st and 2nd columns)
remain nearly identical across all models,
indicating that the duration predictor has accurately captured the characteristic 
of the duration controls in L1 (Japanese).
For rAE, before the modification (3rd column), the mean and SD (especially the latter) are larger 
than those of rNJ (1st column).
This reflects the fact that English, a stress-timed language, exhibits greater variation in 
phoneme duration (especially vowels) compared to Japanese, a mora-timed language. 
After modification (4th column), the values closely align with those for rNJ (1st column), 
suggesting successful duration modeling.
For rJE (5th column), 
the SD is smaller than in rAE (3rd column),
likely due to the influence of their mora-timed L1, 
which constrains variation in phoneme duration even when speaking English. 
The reason for the larger values than rAE after duration modification (4th column)
is that the modification relies only on native Japanese speech.
As a result, the duration patterns resemble those of absolute beginning learners 
with almost no exposure to English sounds,
whereas the real Japanese learners 
have some exposure to English.
Previous studies have shown that as English proficiency increases, L2 speakers exhibit reduced syllable isochrony \cite{MOCHIZUKISUDO1991231, Trofimovich_Baker_2006},
which is consistent with our results.

\subsection{Speech-based analysis}
\begin{figure*}[t]
  \centering
  \hspace*{-2mm}
  \includegraphics[width=\textwidth]{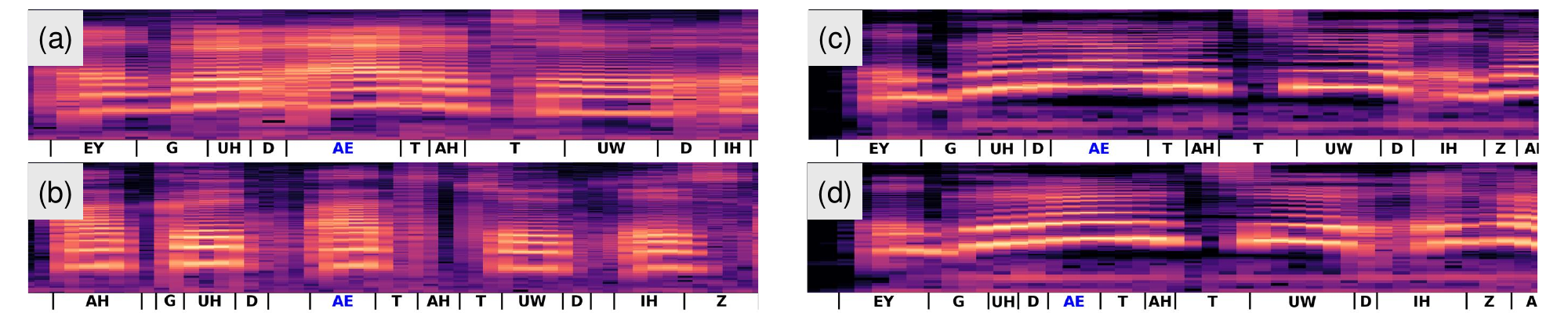}
  \vspace*{-1mm}
  \caption{Forced alignment results of: (a) real American English (USA/M02), (b) real Japanese English (IWA/M01), (c) baseline synthesized Japanese English, (d) duration-modified synthesized Japanese English. (c) and (d) are synthesized using (a) as input speech. The script is ``A good {\color{blue}\textbf{a}}ttitude is unbeatable."}
  \label{fig:fa}
  \vspace*{-3mm}
\end{figure*}
\subsubsection{Degree of accentuation}
To assess the degree of accentuation, we compared each synthesized speech sample with native English.
The evaluation was based on three prosodic features: pitch, intensity, and duration.
For prosody evaluation, we randomly selected four model speakers (two males and two females) 
from rAE for each sentence. 
For each synthesized speech,
we calculated 
the correlation coefficient (Corr.) 
with each of the four model native samples and averaged the results.
For pitch and intensity, Corr. was calculated 
between frame-wise feature sequences,
while for duration, it was computed between frame lengths of phoneme sequence.
The analysis was conducted after aligning the synthesized speech with the model speech 
using PPG-DTW \cite{ppg} and extracting voiced frames via forced alignment. 
The Kaldi-WSJ recipe \cite{Kaldi} was used for PPG extraction and forced alignment.
As a reference, word error rate (WER) was also calculated using a pre-trained Whisper medium model \cite{whisper}.

\begin{table}[tb]
  \centering
  \caption{Correlation coefficients of prosodic features with native English: the smaller the stronger accent} 
  \label{tab:degree}
  \vspace*{-1mm}
  \hspace*{-2mm}
  \resizebox{0.9\columnwidth}{!}{

    \begin{tabular}{llcccc}  \toprule
      \multicolumn{2}{l}{}& \multicolumn{3}{c}{Corr. ($\uparrow$)} &\multirow{2}{*}{WER ($\downarrow$)}\\
      \multicolumn{2}{c}{Model}        & Pitch    & Intensity & Duration & \\ \midrule
      \multirow{3}{*}{sJE(50)} & baseline  &  0.45  & 0.50  & \textbf{0.71} & \textbf{1.18} \\
      & dedup & 0.37 & 0.47  & 0.52  & 1.25\\
      & dur-mod &  0.43 & 0.50  & 0.54 & 1.26\\ \midrule
      \multirow{3}{*}{sJE(200)} & baseline  & 0.43 & 0.56  & \textbf{0.74} & \textbf{0.88} \\
      & dedup &  0.42  & 0.54  & 0.63 & 1.08\\
      & dur-mod &  0.44 & 0.55  & 0.62 & 1.05\\ \midrule
      \multirow{3}{*}{sJE(1000)} & baseline & 0.47 & 0.56  & \textbf{0.75} & \textbf{0.73}\\
      & dedup &  0.45 &  0.56 & 0.65 & 0.95\\
      & dur-mod  &  0.47 & 0.55  & 0.66 & 0.87\\ \midrule
      rJE &  & 0.47 &  0.49  & 0.51& 0.15 \\ \midrule
      rAE &   &  \textbf{0.69}  & \textbf{0.68}  & \textbf{0.78} & \textbf{0.03}\\ \bottomrule

    \end{tabular}
  }
  \vspace*{-3mm}
\end{table}

Table \ref{tab:degree} shows the results. 
As a reference, the results for rJE and rAE are also shown, as well as synthesized Japanese English (sJE).
The values of rAE can be seen as an upper bound for the Corr. with native English.
When comparing models with the same number of units, the WER for the deduplication-only model (dedup) 
and the duration-modification model (dur-mod) are higher than that of the baseline model.
This is due to the decrease in intelligibility caused by the durational accents introduced by the duration modification module.
In fact, the correlation shows that the duration of the baseline model's output is closely matches 
that of native English,
while the outputs of the other two models demonstrate clear deviations from native English.
While the duration modification successfully added accents on duration,
the degree of accentuation for pitch and intensity remained almost unchanged.
Given that the correlation values for pitch and intensity are almost the same as those for rJE,
this outcome can be considered as a successful result.

For the number of units, 
there is a general trend that as the number of units increases, 
the similarity to native English improves for all prosodic features.
This indicates that the amount of the prosodic information encoded in the units
increases with the number of units.
However, the trend is somewhat weak, especially for pitch,
and the relationship with the language used for S2u training remains unclear.
Future work will focus on investigating the type of information (segmental, prosodic, or acoustic) 
encoded in the units, and how this is influenced by the tokenization method, 
including the number of units and the language used for training.



\subsubsection{Tendency of durational accents seen in Japanese English} 
\begin{table}[tb]
  \centering
  \caption{Average duration of stressed/unstressed vowels}
  \label{tab:duration}
  \vspace*{-1mm}
  \resizebox{0.9\columnwidth}{!}{
    \begin{tabular}{llrrr}  \toprule
      \multicolumn{2}{l}{Model}&Stressed [ms] & Unstressed [ms] & Ratio \\ \midrule
      \multirow{3}{*}{sJE(50)} & baseline & 108 & 72 & 1.48 \\
      & dedup & 109 & 80 & 1.37 \\
      & dur-mod & 96 & 72 & \textbf{1.33} \\ \midrule
      \multirow{3}{*}{sJE(200)} & baseline & 109 & 74 & 1.47 \\
      & dedup & 107 & 73 & 1.46 \\
      & dur-mod & 91 & 63 & \textbf{1.43} \\ \midrule
      \multirow{3}{*}{sJE(1000)} & baseline & 107 & 72 & 1.50 \\
      & dedup & 103 & 71 & 1.47 \\
      & dur-mod & 88 & 62 & \textbf{1.41} \\ \midrule
      rJE && 135 & 115 & 1.18 \\ \midrule
      rAE & &101 & 70 & 1.43 \\ \bottomrule

    \end{tabular}
  }
  \vspace*{-3mm}
\end{table}

While the previous section showed that the duration modification module
successfully added accents on duration to the output speech,
it is still unclear whether these accents are similar to those found in real Japanese English.
In this section, we examine a typical characteristic of accents commonly seen in real Japanese English 
to assess whether they are also present in the synthesized speech.
Japanese is a mora-timed language, where each mora tends to have nearly the same duration \cite{morajp}. 
Similarly,
Japanese-accented English often exhibits a syllable-timed rhythm \cite{riney1993japanese, stress}, 
with all vowels pronounced with almost the same duration.
In contrast, English is originally a stress-timed language, 
where stresses occur at regular intervals, 
and unstressed vowels between stresses are generally shortened \cite{DAUER198351}.

Figure \ref{fig:fa} shows an example of the forced alignment results for 
rAE, rJE, baseline sJE(1000), and dur-mod sJE(1000)
for the sentence ``A good attitude is unbeatable''.
The primary stress of the word ``attitude'' is placed on the first syllable. 
In rAE, the duration of the stressed vowel ``a'' is longer than that of the unstressed vowels, 
while in rJE the length of the vowel is almost the same as other vowels.
In the baseline sJE, the vowel durations are almost the same as that of input rAE,
failing to replicate the durational accents of Japanese English.
On the other hand, in the dur-mod sJE, the duration of the ``a'' is shortened
and becomes almost the same as that of the other vowels.

Table \ref{tab:duration} shows the overall average 
of phoneme durations for stressed and unstressed vowels, along with their ratio.
As expected, the ratio for rJE is lower than that of rAE.
In the synthesized speech, for all the number of units, 
the duration modification model has the lowest ratio, which is closer to that of rJE.
These results
indicate that the duration modification module successfully 
replicates the isochronization of vowels, a typical feature of Japanese English.
\subsection{Subjective evaluation}
\begin{figure}[t]
  \centering
  \hspace*{-3mm}
  \includegraphics[width=1.1\columnwidth]{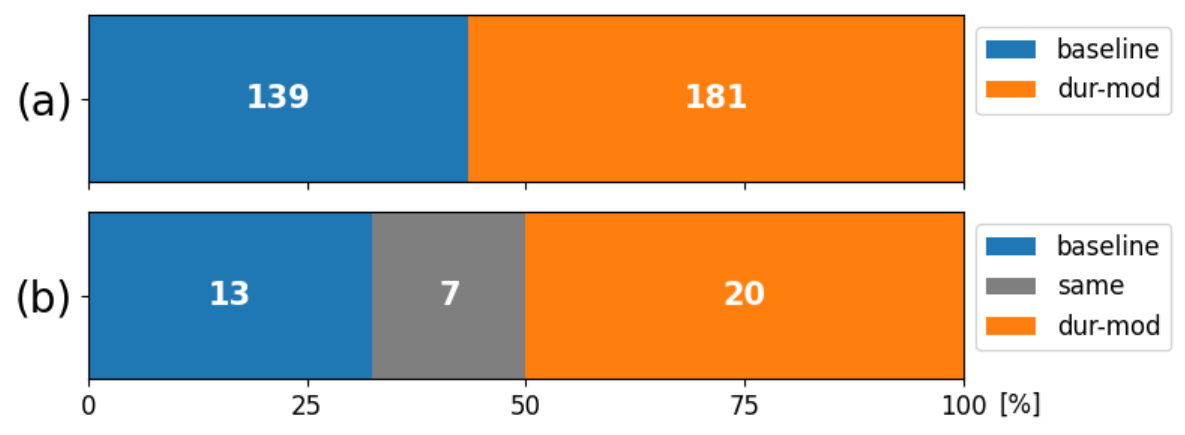}
  \vspace*{-6mm}
  \caption{Preference results as Japanese-accented English: a) total response counts, b) majority votes per sentence.}
  \label{fig:eval}
  \vspace*{-4mm}
\end{figure}
For the subjective evaluation, eight Japanese teachers of English 
with at least five years of experience teaching English to Japanese students 
were recruited. 
They were presented with Japanese English speech synthesized 
using both the baseline and proposed methods, 
and were asked to determine which one was closer to 
the characteristics of real Japanese pronunciation. 
Using models with the cluster size of 1000,
we compared speech synthesized from the same American English speech 
using the baseline and dur-mod methods. 
To focus the evaluation on differences in accent characteristics 
rather than intelligibility, the script was displayed alongside the audio. 
Speech pairs that both were fully recognized by the Whisper medium model
were pre-selected, and 40 pairs were randomly chosen from these.
The same test set was used for all evaluators.

The results are shown in Figure \ref{fig:eval}.
The total number of responses for each method and
the results of majority votes among the evaluators for each sentence are shown.
For the majority vote, when both methods got the same number of votes,
the sentence was classified as ``same."
Both results showed that the proposed method was preferred over the baseline method
as a more natural Japanese-accented English.
This indicates that the proposed method successfully reproduced the characteristics of 
Japanese English pronunciation more accurately.
The objective evaluation confirmed that the proposed method appropriately
added accents on duration,
and now it is also confirmed that the accents as a whole
are perceived as more natural
by human experts on Japanese-accented English.

\section{Conclusions and future work}
In this paper, we introduced a duration-modification module for foreign accent simulation 
with discrete token-based resynthesis using only native speech corpora.
The evaluation results showed that the module successfully added natural accents to the phoneme duration.
Especially, the isochronization of vowels, a typical feature of Japanese English, was successfully reproduced.
Subjective evaluation by Japanese English teachers also confirmed 
that the synthesized speech using the
proposed method had more natural Japanese accents.

To further validate the effectiveness of the proposed method, 
we will examine whether the output speech enhance
the robustness of both human and machine listeners to real accented speech. 
Also, we need to investigate 
what kind of information is encoded in the units 
and explore the way of tokenization for better accentuation,
while using only native speech corpora.

\section{Acknowledgements} 
This work was supported by AIST KAKUSEI project (FY2024).

\bibliographystyle{IEEEtran}
\bibliography{references}

\end{document}